# Towards Intrinsic Charge Transport in Monolayer Molybdenum Disulfide by Defect and Interface Engineering

Zhihao Yu[1,†], Yiming Pan[2,†], Yuting Shen[3], Zilu Wang[4], Zhun-Yong Ong[5], Tao Xu[3], Run Xin[1], Lijia Pan[1], Baigeng Wang[2], Litao Sun[3,*], Jinlan Wang[4], Gang Zhang[5], Yong Wei Zhang[5], Yi Shi[1,*] & Xinran Wang[1,*]

[1]*National Laboratory of Solid State Microstructures, School of Electronic Science and Engineering, National Center of Microstructures and Quantum Manipulation, Nanjing University, Nanjing 210093, P. R. China*

[2]*School of Physics, Nanjing University, Nanjing 210093, P. R. China*

[3]*SEU-FEI Nano-Pico Center, Key Laboratory of MEMS of Ministry of Education, Southeast University, Nanjing, 210096, P. R. China*

[4]*Department of Physics, Southeast University, Nanjing 211189, P. R. China*

[5]*Institute of High Performance Computing, 1 Fusionopolis Way, 138632, Singapore*

† These authors contribute equally to this work.

* Correspondence should be addressed to X. W. (xrwang@nju.edu.cn) or Y. S. (yshi@nju.edu.cn) or L. S. (slt@seu.edu.cn).

**Abstract**

Molybdenum disulfide is considered as one of the most promising two-dimensional semiconductors for electronic and optoelectronic device applications. So far, the charge transport in monolayer molybdenum disulfide is dominated by extrinsic factors such as charged impurities, structural defects and traps, leading to much lower mobility than the intrinsic limit. Here, we develop a facile low-temperature thiol chemistry to repair the sulfur vacancies and improve the interface, resulting in significant reduction of the charged impurities and traps. High mobility greater than 80cm$^2$ V$^{-1}$ s$^{-1}$ is achieved in backgated monolayer molybdenum disulfide field-effect transistors at room temperature. Furthermore, we develop a theoretical model to quantitatively extract the key microscopic quantities that control the transistor performances, including the density of charged impurities, short-range defects and traps. Our combined experimental and theoretical study provides a clear path towards intrinsic charge transport in two-dimensional dichalcogenides for future high-performance device applications.

Despite its great promise as a two-dimensional (2D) channel material for logic transistors[1,2], integrate circuits[3,4] and photodetectors[5,6], the charge transport in monolayer of molybdenum disulfide (MoS$_2$) is still far away from the intrinsic limit. Theoretically, the phonon-limited mobility is ~410cm$^2$V$^{-1}$s$^{-1}$ at room temperature, and is weakly dependent on carrier density until ~10$^{13}$cm$^{-2}$, where electron-electron scattering starts to play important roles[7]. Experimentally however, very different behavior is observed regardless of the sample preparation method. Firstly, the electron mobility of backgated monolayer MoS$_2$ devices is limited to ~40cm$^2$V$^{-1}$s$^{-1}$ at room temperature, an order of magnitude lower than the phonon limit [2, 8-10]. Secondly, the mobility is found to increase with carrier density even beyond 10$^{13}$cm$^{-2}$ (Ref. 2, 9). Thirdly, at low carrier density, the charge transport is dominated by hopping mechanism[11, 12]. Fourthly, a metal-insulator transition (MIT) is observed at high carrier densities on the order of 10$^{13}$cm$^{-2}$, but the underlying mechanism remains debated[2, 8, 9]. These observations clearly point to the existence of extrinsic factors that dominate charge transport in monolayer MoS$_2$. Charged impurities (CI)[2, 13], short-range defects[11] and localized states[11, 14, 15], among others, are believed to strongly influence electron transport. However, a comprehensive physical picture that explains all of above-mentioned transport phenomena and provides quantitative and microscopic insights into the impurities is still lacking.

Due to its atomic thickness, electrons in monolayer MoS$_2$ are more susceptible to impurities both inside MoS$_2$ and at the dielectric interface. Therefore, the engineering of defects and interface represents a logical route to further improve MoS$_2$ device performance. Recently, the use of thiol-terminated SiO$_2$ (Ref. 16), boron nitride[17], and suspended structure[18] were found to improve the mobility of monolayer MoS$_2$ devices by several folds, showing the importance of interface. However, their mobility values are still much lower than the intrinsic limit, indicating that other sources of impurities, most likely inside $MoS_2$, are still present.

In this work, we show that sulfur vacancies (SVs), which are the main type of intrinsic defects in $MoS_2$ (Ref. 11), can be effectively repaired by (3-mercaptopropyl)trimethoxysilane (MPS) under mild annealing, resulting in significant reduction of CI and short-range scattering. Monolayer $MoS_2$ with both sides treated by MPS exhibits a record-high mobility greater than $80 cm^2V^{-1}s^{-1}$ ($300 cm^2V^{-1}s^{-1}$) at room temperature (low temperature), much higher than untreated samples. In addition, we show that MIT in $MoS_2$ is due to localized trap states, which can be modulated by improving the sample and interface quality. A theoretical model that takes into account the major scattering sources (phonon, CI and short-range defects) and localized charge traps is developed to quantitatively understand the scaling of mobility, conductivity and MIT in monolayer $MoS_2$. By fitting the experimental data, we are able to extract key microscopic quantities including the density of CI, short-range defects and charge traps, as well as derive a transport phase diagram for $MoS_2$. The quantitative information also allows us to discuss the possible origins of these extrinsic factors, which serve as the basis for further device optimization.

**Repair of sulfur vacancies by thiol chemistry**

The monolayer $MoS_2$ samples studied here are obtained by mechanical exfoliation from bulk crystals (Supplementary Fig. 1). A high density of SV exists in as-exfoliated $MoS_2$ as demonstrated in earlier works [11, 19]. These defects, which can act as catalytic sites for hydrodesulfurization reactions [20, 21], are chemically reactive. Therefore, it is possible to repair the SV by thiol chemistry. Here, we choose a specific molecule MPS (Fig. 1 inset) for two reasons. (a) The S-C bond in MPS is weaker than other thiol molecules like dodecanethiol due to the acidic nature of $CH_3$-O- groups, leading to a low energy barrier for the reaction[22]. (b) The trimethoxysilane groups in MPS react with the $SiO_2$ substrate to form a self-assembled monolayer[23] (SAM, Supplementary Fig. 2). The SAM layer can effectively passivate the $MoS_2/SiO_2$ interface, while the outstanding thiol group can also repair the SV on the bottom side of $MoS_2$, which is otherwise difficult to access. This unique property of MPS allows us to systematically compare three types of $MoS_2$ samples: as-exfoliated on $SiO_2$, top-side (TS) treated on $SiO_2$, and double-side (DS) treated. For the MPS treatment, we used a liquid-phase process[22], followed by 350 °C annealing in forming gas to repair the SV (see Methods and Supplementary Fig. 3 for details of sample preparation and characterization).

We first study the reaction kinetics of SV and MPS by density functional theory (Fig. 1). The reaction can be described by $HS(CH_2)_3Si(OCH_3)_3 + SV \rightarrow CH_3(CH_2)_2Si(OCH_3)_3$ (Supplementary Fig. 4, Supplementary Movie 1), which is exothermic with the enthalpy change of -333.3kJ $mol^{-1}$ of MPS. The reaction process comprises of two steps with an energy barrier of 0.51eV and 0.22eV, respectively. In the first step (Supplementary Fig. 5), MPS chemically absorbs on to an SV through the sulfur atom, and then it cleaves the S-H bond and forms a thiolate surface intermediate. The dissociated H atom is bonded to a neighboring S atom. The second step involves cleavage of the S-C bond and hydrogenation of the thiolate intermediate to form the final product trimethoxy(propyl)silane (Supplementary Fig. 6). We note that the two-step process agrees well with earlier thiol absorption experiments on defective $MoS_2$[20, 22]. The relatively low energy barriers also facilitate the reaction to happen at low temperature.

To quantify the effect of MPS treatment on sample quality, we perform aberration-corrected transmission electron microscopy (TEM) on as-exfoliated and TS-treated monolayer MoS$_2$. Fig. 2 compares high-resolution TEM images of a typical as-exfoliated and TS-treated sample. The SVs can be clearly distinguished by analyzing the intensity profile[11] as shown in Supplementary Fig. 9. Statistical analysis (from more than 15 areas in each case) indicates that the density of SV is reduced from ~6.5x10$^{13}$cm$^{-2}$ for the as-exfoliated samples to ~1.6x10$^{13}$cm$^{-2}$ for the TS-treated samples (Supplementary Fig. 9c, d). Due to the difficulty in making TEM samples, we are not able to characterize the DS-treated MoS$_2$, where further reduction of SV is expected. During the course of TEM experiment, we have paid great attention to prevent the knock-on damage and lattice reconstruction caused by electron beam irradiation[24] by limiting the exposure time to less than 30s and the current density to below 10$^6$ e nm$^{-2}$S$^{-1}$ (Ref. 11). The SV generation rate induced by electron irradiation under our experimental conditions was very low, ~5.6x10$^{10}$cm$^{-2}$S$^{-1}$ (Ref. 11). Therefore the SVs in Fig. 2 are believed to be intrinsic rather than induced by electron irradiation.

**Electrical transport properties**

Next, we systematically investigated the effect of MPS treatment on electrical transport properties of MoS$_2$. We fabricated backgated field-effect transistors (FETs) on as-exfoliated, TS-treated and DS-treated monolayer MoS$_2$ samples and carried out electrical measurements in high vacuum with standard four-probe technique, unless otherwise stated (see Methods for details of device fabrication and measurement). All the devices exhibited very small hysteresis at room temperature, which became even smaller as they were cooled down (Supplementary Fig. 10). Therefore, in the following we only present electrical data from the forward sweep.

Fig. 3a shows the four-probe conductivity $\sigma = \dfrac{I_{ds}}{\Delta V} \dfrac{L}{W}$ as a function of backgate voltage $V_g$ for three representative devices at room temperature (300K), where $I_{ds}$ is the source-drain current; $\Delta V$, $L$ and $W$ are the voltage difference, distance, and sample width between the two voltage probes, respectively. The MPS-treated samples show improved $\sigma$ compared to the as-exfoliated one, indicating higher sample and interface quality. At carrier density $n=C_gV_g=7.1\times10^{12}cm^{-2}$ ($C_g=11.6nFcm^{-2}$ is the gate capacitance for 300nm $SiO_2$ dielectrics), the DS-treated sample shows $\sigma=1.52e^2h^{-1}$ and field-effect mobility $\mu=\dfrac{d\sigma}{C_g dV_g}=81cm^2V^{-1}s^{-1}$. To our best knowledge, this is the highest room-temperature field-effect mobility reported so far for monolayer $MoS_2$ regardless of the device geometry [2, 8-10].

To gain further insights into the charge transport physics, we performed variable-temperature electrical measurements down to 10K. Surprisingly, the three types of devices exhibit very different behavior (Fig. 3, Supplementary Fig. 11). For the as-exfoliated sample, $\sigma$ monotonically decreases during cooling over the entire range of $n$, indicating an insulating behavior (Fig. 3d, Supplementary Fig. 11a). For the DS-treated sample, the transfer curves all intersect near $V_g=80V$ (corresponding to $n\sim5.7\times10^{12}cm^{-2}$, Supplementary Fig. 11c), which is a signature of MIT. Metallic and insulating behaviors are observed down to the base temperature for $n>6.6\times10^{12}cm^{-2}$ and $n<3.5\times10^{12}cm^{-2}$ respectively. At intermediate $n$, the $\sigma$-$T$ characteristics are not monotonic and MIT occurs within our experimental temperature range (solid symbols in Fig. 3f). For the TS-treated sample, MIT is observed for $n>4.7\times10^{12}cm^{-2}$ (solid symbols in Fig. 3e), while insulating behavior is always observed at low temperatures (Fig. 3e, Supplementary Fig. 11b). As we shall see later, these distinct transport behaviors precisely reflect the quality of the $MoS_2$ and interface.

The scaling behavior of $\mu$ is also very different for the three samples (Fig. 3b-c). The as-exfoliated sample shows the lowest $\mu$ among the three samples. The $\mu$-$T$ curve is not monotonic with a peak value of 31cm$^2$V$^{-1}$s$^{-1}$ near 175K. On the other hand, $\mu$ monotonically increases for the MPS-treated samples upon cooling. For $T$>100K, $\mu \sim T^{-\gamma}$, where $\gamma$=0.72 and 0.64 for the DS-treated and TS-treated sample respectively. For $T$<100K, $\mu$ gradually saturates. At $T$=10K, $\mu$=320 cm$^2$V$^{-1}$s$^{-1}$ for the DS-treated sample, which is ~3 (22) times higher than the TS-treated (as-exfoliated) one.

We stress that all the above-mentioned transport phenomena are reproducible among different samples. In Supplementary Fig. 12, we show three additional sets of data for two-terminal devices. Although the mobility is lower than their corresponding four-terminal counterparts due to contact resistance, all the key transport properties, including MIT, scaling of mobility and conductivity, are qualitatively reproduced.

**Theoretical modeling of charge transport**

Although some theories have been proposed to explain the mobility of monolayer MoS$_2$[13, 25], a more complete physical model remains to be developed to fully understand the charge transport. In particular, the model should establish the correlation between different transport regimes and the underlying microscopic scattering mechanisms, provide quantitative information about the samples, and project device performance based on realistic parameters.

We start by calculating the mobility of monolayer MoS$_2$. According to Matthiessen's rule, the mobility for free carriers is expressed as

$$\mu_0(n,T)^{-1} = \mu_{ph}(T)^{-1} + \mu_{CI}(n,T)^{-1} + \mu_{sr}^{-1} \qquad (1)$$

where $\mu_{ph}$, $\mu_{CI}$ and $\mu_{sr}$ are mobility limited by phonons, CI, and short-range scatterings, respectively (the calculation of each term is described in Methods). The incorporation of $\mu_{sr}$ in our model is motivated by (i) TEM characterization that clearly shows the existence of short-range SV defects and (ii) the saturation of $\mu$ at low temperatures. We do not consider surface optical phonon scattering because the relatively high energy of the $SiO_2$ phonon modes makes them irrelevant to the low-field mobility phenomenon considered here[25].

The experimentally measured "effective" field-effect mobility $\mu$ is not equal to the free-carrier mobility $\mu_0$, due to the presence of charge traps that limits the free carrier population. Recently, localized trap states within the bandgap have been observed in both exfoliated and CVD monolayer $MoS_2$ (Ref. 11, 14, 15). The impurity band from these trap states can introduce a mobility edge that strongly affects the charge transport [26]. In a simple picture [27], transport is carried only by electrons in the extended states, *i.e.* states above the mobility edge. This model, which does not account for the hopping between the localized states, has been very successful in modeling organic FETs [27, 28]. Here we adopt the same model to extract important physical quantities such as the density of CI ($N_i$) and trap states ($N_{tr}$), while avoiding the complexity of dealing with the energy-dependent mobility and percolation effects in hopping transport[27] (the detailed calculation is described in Methods). The model is further justified by its excellent agreement with the experimental data over a broad range of temperature and carrier density.

**Discussion**

With the above model, we can now quantitatively understand the obtained experimental data. The solid lines in Fig. 3 are the best fitting results using the parameters listed in Table 1. Remarkably, the scaling of mobility and conductivity with temperature and carrier density is well reproduced with a single set of parameters, suggesting that our model captures the essential physics. At low temperature and low carrier density, the calculated $\sigma$ and $\mu$ are lower than experiments (Fig. 3b, d-f), presumably due to the omission of hopping transport in our model. Hence, the discrepancy is the largest for the as-exfoliated sample (Fig. 3b, Fig. 3d) because $N_{tr}$ is the highest.

The fitting parameters in Table 1 give considerable insights into the microscopic origin of impurities in MoS$_2$. In all of our samples, $\mu$ is much lower than $\mu_{ph}$, indicating that phonon scattering does not play a significant role. Rather, the mobility is largely limited by CI and short-range scatterings. We notice that $N_i$ and $\mu_{sr}$ are reduced by MPS treatment and are correlated for each sample, suggesting that CI partially shares the microscopic origin with short-range defects, most likely SV. CI also partially comes from the interface, as the DS-treated sample has much lower $N_i$ than TS-treated one. For the DS-treated sample, $N_i$ becomes comparable to that of SiO$_2$ substrate (0.24-2.7x10$^{11}$cm$^{-2}$) [29-31], indicating that a large portion of SV is repaired. $N_{tr}$ is also partially due to SVs as it can be reduced by MPS treatment. However, $N_{tr}$ is an order of magnitude higher than $N_i$ for all the samples, pointing to contributions from additional sources. This could be due to the ambient species absorbed between MoS$_2$ and SiO$_2$ during exfoliation, which act as charge traps as in the case of graphene [29, 31, 32].

For the MPS-treated samples, the high-temperature mobility is limited by CI, leading to the usual $T^{-\gamma}$ scaling behavior. Ref. 13 predicts $\gamma$~1 for the range of carrier densities studied here. However, both short-range scattering (temperature independent) and thermal excitation from charge traps (which leads to the opposite trend since the density of electrons in the extended states $n_c$ increases with temperature) can decrease the effective $\gamma$ as observed here. Therefore, one cannot reliably infer the scattering mechanism solely by analyzing $\gamma$. When the charge traps become dominant (which usually happens at low temperature and low carrier density), the mobility even exhibits an insulating behavior as commonly observed in backgated devices [2, 11, 12] and in the as-exfoliated sample here (Fig. 3b).

The mobility increases with carrier density for all three samples from the combined effect of CI and charge traps (Fig. 3c). Below a threshold carrier density (equivalent to mobility edge), $\mu$ becomes negligible due to the localized nature of electrons in charge traps. The mobility edge is found to decrease with temperature (Supplementary Fig. 13) because of the thermal excitation of electrons to the extended states. At $T$=80K, the threshold density is about $2 \times 10^{12}$, $3 \times 10^{12}$ and $5 \times 10^{12}$ cm$^{-2}$ for the DS-treated, TS-treated and as-exfoliated sample respectively (Fig. 3c). These values are on the same order of magnitude with $N_{tr}$, further supporting the charge trap model.

Finally, we can also understand MIT in the framework of charge traps. Strong electron-electron correlation in 2D electron gas has been proposed to explain the MIT in monolayer MoS$_2$ which gives a universal threshold density $n_{MIT} \sim 10^{13}$cm$^{-2}$ (Ref. 2). In our experiments, however, $n_{MIT}$ is apparently dependent on sample quality (Fig. 3). The MIT can be intuitively understood from Eq. 6 (Methods). When $n \ll N_{tr}$, the density of conducting electrons in the extended states ($n_c$) is exponentially dependent on temperature due to thermal activation, the temperature dependence of $\sigma$ is dominated by $n_c$, showing thermally activated insulating behavior [2, 11, 12, 14]. For even smaller $n$ and $T$, transport is dominated by variable-range hopping because $n_c$ can be ignored [11, 14] (Fig. 4a). When $n \gg N_{tr}$, the Fermi level is far above the mobility edge and $n_c$ is independent of temperature. Thus, the temperature dependence of $\sigma$ is dominated by $\mu_0$, showing metallic behavior. Therefore, MIT occurs when $n \approx N_{tr}$. Using the trap and CI parameters of the DS-treated sample in Table 1, we numerically calculated $\sigma(n,T)$ (Eq. 6) and obtained the $n$ and $T$ of each critical point as in Fig. 3f. Fig. 4a shows the calculated transport phase diagram with metallic and insulating regions. Excellent agreement with experiment is achieved without any fitting

12parameters, showing the consistency of our model. From Fig. 4a, $n_{MIT}$ slowly increases with decreasing temperature, and converges to $n_0$ at $T$=0K. When $n>n_0$, metallic behavior is always expected. When $n<n_0$, MIT is always expected at finite temperature. Under realistic trap and CI parameters, $n_0$ is linearly proportional to, but slightly higher than $N_{tr}$ (Fig. 4b, Supplementary Fig. 14). Therefore, $n_0$ can be used as a rough estimate of $N_{tr}$. After careful comparison with literature [2, 9, 14, 33], we find that the $N_{tr}$ of our MPS-treated samples is indeed the lowest (Fig. 4b), consistent with the lowest observed $n_0$.

From the above discussion, it is clear that the transport in current state-of-the-art monolayer $MoS_2$ samples is still limited by charge traps, CI and short-range defects. To reach the real potential of monolayer $MoS_2$ in high-performance devices, continuous improvement of sample and interface quality is still needed. In Fig. 4c, we project the room-temperature mobility as a function of $N_{tr}$ in the ideal case, i.e. without CI and short-range scattering. The mobility at low carrier density is rapidly degraded by even a small $N_{tr}$. At a modest $n=7\times10^{12}cm^{-2}$, $N_{tr}$ has to be lower than $8.8\times10^{11}cm^{-2}$, which is ~6 times lower than our DS-treated device, in order to achieve mobility greater than $400 cm^2V^{-1}s^{-1}$.

In conclusion, we have shown that thiol chemistry is an effective approach to engineer the defects and interface in monolayer $MoS_2$ towards intrinsic charge transport. A physical model that includes charge traps and major scattering sources has been developed to comprehensively describe the charge transport in $MoS_2$ and to quantify the density of CI and charge traps in the samples. We believe that our model captures the essential charge transport physics for monolayer $MoS_2$ and can be readily extended to other 2D semiconductors [33-35].

## Methods

### DFT calculations

DFT calculations were performed using the Vienna *ab initio* simulation package (VASP)[36, 37]. Projector-augmented-wave (PAW)[38] potentials were used to describe ion-electron interactions, and the exchange correlation potential was represented by the local density approximation (LDA)[39]. We used the climbing-image nudged elastic band (cNEB) method to locate the minimum energy paths and the transition states[40]. The defective $MoS_2$ sheet was modeled by a $4\times4$ supercell of $MoS_2$ with a single SV. A k-point sampling of $5\times5\times1$ was used for all the calculations.

### $MoS_2$ sample preparation and MPS treatment

In this work, we exfoliated monolayer $MoS_2$ from natural bulk flakes (SPI Supplies). The as-exfoliated samples were directly exfoliated on 300nm $SiO_2$/Si substrate. Prior to device fabrication, the samples were annealed in a mixture of $H_2$/Ar at 350 ℃ to remove organic residue.

For the TS-treated samples, we first exfoliated monolayer $MoS_2$ on 300nm $SiO_2$/Si substrate, followed by annealing in a mixture of $H_2$/Ar at 350 ℃ to remove organic residue. The sample was then dipped in a fresh solution of 1/15 (v/v) MPS/dichloromethane for 24 hours in a dry glove box to grow MPS on $MoS_2$. After growth, the sample was taken out, rinsed thoroughly with dichloromethane and IPA and blown dry with $N_2$. Finally, the sample was annealed in a mixture of $H_2$/Ar at 350 ℃ for 20 minutes to finish the MPS treatment and remove the extra MPS on $MoS_2$ (Supplementary Fig. 3).

For DS-treated samples, the 300nm $SiO_2$/Si substrate was first subject to a 30-minute UV/ozone treatment to enhance surface hydrophilicity. The substrate was then dipped in a 10% (v/v) MPS/dichloromethane solution for 12 hours in a dry glove box to grow MPS SAM. After the SAM growth, the substrate was sonicated in dichloromethane followed by thorough rinsing with dichloromethane and IPA and drying with $N_2$ (Supplementary Fig. 3). We then exfoliated monolayer $MoS_2$ on the SAM treated $SiO_2$/Si substrate. After exfoliation, the sample was dipped in a fresh solution of 1/15 (v/v) MPS/dichloromethane for 24 hours in a dry glove box to grow MPS on the top side of $MoS_2$, followed by thorough rinsing. Finally, the sample was annealed in a mixture of $H_2$/Ar at 350 ℃ for 20 minutes to finish the MPS treatment.

**Device fabrication and electrical measurement**

We used standard photolithography to pattern the electrodes of $MoS_2$ backgated FETs on 300nm $SiO_2$/Si substrates. Ti/Pd (20nm/20nm) was electron beam evaporated as the contact metal for source, drain and voltage probes, followed by lift-off. All devices were annealed at 350 ℃ in Ar to improve contacts. Electrical measurements were carried out by an Agilent B1500 semiconductor parameter analyzer in a close-cycle cryogenic probe station with base pressure $\sim 10^{-5}$ Torr. Prior to electrical measurements, a vacuum annealing at base pressure was performed to remove adsorbates and improve device performances[41].

**TEM characterizations**

To prepare TEM samples, monolayer $MoS_2$ was first exfoliated on 300nm $SiO_2$/Si substrate. We then align-transferred the sample to a copper TEM grid by the PMMA-based transfer method described in Ref. 11.

The high-resolution TEM images were recorded in an image aberration-corrected

TEM (FEI Titan 80–300 at 80 kV) equipped with a charge-coupled device camera (GatanUltraScanTM 1000). The exposure time was 1s. Before recording, we set the third-order spherical aberration in the range of 1–6 μm and then we recorded the images under slightly underfocused. In order to minimize irradiation damage and prevent new SVs generating during normal imaging [24], we limited the total exposure time to less than 30 s and the current density to lower than $10^6$ e/nm²/S. A band-pass filter to the Fast Fourier Transformation of the original images was applied to enhance the visibility of the defects.

**Theoretical modeling**

The phonon-limited mobility is numerically calculated according to Ref. 7:

$$\mu_{ph}(T) = \begin{cases} \alpha_1 T^{-1}, T < 100K \\ \alpha_2 T^{-1.69}, T \geq 100K \end{cases} \quad (2)$$

where the coefficients $\alpha_1 = 2.625 \times 10^5 cm^2 KV^{-1}s^{-1}$ and $\alpha_2 = 6.297 \times 10^6 cm^2 KV^{-1}s^{-1}$ satisfy the boundary condition that $\mu_{ph}(300K) = 410 cm^2 V^{-1}s^{-1}$ and that $\mu_{ph}(T)$ is continuous at 100K. The transition at 100K is due to the different phonon modes that limit the mobility [7].

The CI-limited mobility is calculated by the model described in Ref. 13:

$$\mu_{CI}(n,T) = \frac{e}{\pi n \hbar^2 k_B T} \int_0^\infty f(E)[1-f(E)]\Gamma_{imp}(E)^{-1} E dE \quad (3)$$

where $f(E)$ is the equilibrium Fermi-Dirac distribution function and $\Gamma_{imp}$ is the electron momentum relaxation rate given by

$$\Gamma_{imp} = \frac{N_i}{2\pi\hbar} \int dk' \left|\phi^{scr}_{|k-k'|}\right|^2 (1-\cos\theta_{kk'})\delta(E_k - E_{k'}) \quad (4)$$

where $\phi^{scr}_{|q|}$ is the screened scattering potential for a single CI. The expression and derivation for $\phi_{|q|}^{scr}$ can be found in Ref. 13. Physically speaking, $\phi_{|q|}^{scr}$ includes the effect of temperature-dependent charge polarizability on the electron screening of the charged impurity, which dominates the temperature dependence of the mobility at low and room temperature. The expression also implies that $\mu_{CI}$ is inversely proportional to $N_i$. Hence, from the fitting of the $\mu_{CI}$ to the total effective mobility data, we are able to quantify the $N_i$ within the MoS$_2$. As an example, Supplementary Fig. 15 plots $\mu_{CI}$ as a function of $T$ under $N_i=10^{12}$cm$^{-2}$.

The short-range-scattering-limited mobility $\mu_{sr}$ is a constant fitting parameter that does not depend on temperature or carrier density[8].

To model the effect of charge traps, we assume the charge traps are uniformly distributed within $\Delta E_{tr}$ below the conduction band edge (Supplementary Fig. 16). We only consider shallow traps because the density of deep traps is at least an order of magnitude lower[14], and that deep traps do not affect mobility and conductivity[27]. The Fermi energy $E_F(n, T)$ is determined by

$$n = C_g V_g = \int_0^{+\infty} N_0 \frac{1}{e^{(E-E_F)/k_B T}+1} dE + \int_{-\Delta E_{tr}}^{0} \frac{N_{tr}}{\Delta E_{tr}} \frac{1}{e^{(E-E_F)/k_B T}+1} dE \quad (5)$$

where $N_0 = \dfrac{2m^*}{\pi \hbar^2} = 3.8 \times 10^{14}$ eV$^{-1}$cm$^{-2}$ is the density of states in the conduction band, and $m^* = 0.45 m_e$ is the conduction band effective mass[42]. The density of conducting electrons in the extended states is

$$n_c(n,T) = \int_0^{+\infty} N_0 \frac{1}{e^{(E-E_F)/k_B T}+1} dE \quad (6)$$

The conductivity is calculated by

$$\sigma(n,T) = e n_c(n,T) \mu_0(n,T) \quad (7)$$

and the "effective" mobility is given by

$$\mu(n,T) = \mu_0(n,T) \frac{\partial n_c(n,T)}{\partial n} \tag{8}$$

**END NOTES**

**Acknowledgements.** This work was supported in part by Chinese National Key Fundamental Research Project 2013CBA01604, 2011CB922103, 2011CB707601, 2010CB923401, 2011CB302004; National Natural Science Foundation of China

61325020, 61261160499, 11274154, 61229401, 21173040, 21373045, 113279028, 61274114; National Science and Technology Major Project 2011ZX02707; Natural Science Foundation of Jiangsu Province BK2012302, BK20130055, BK20130016, BK2012024. J. W. acknowledges the computational resources provided by SEU and National Supercomputing Center in Tianjin.

**Author Contributions.** X. W and Y. S. conceived and supervised the project. Z. Y. and R. X. carried out sample preparation, device fabrication, electrical measurements and data analysis. Y. P., Z.-Y. O, B. W., G. Z. and Y. W. Z. performed modeling of charge transport. Y. S., T. X. and L. S. performed TEM characterizations and analysis. Z. W. and J. W. performed DFT calculations. L. P. devised the MPS treatment method. X. W., Y. S., L. S. and J. W. co-wrote the paper with all authors contributing to discussion and preparation of the manuscript.

**Competing financial interests statement.** The authors declare no competing financial interest.

**Figure Captions**

**Figure 1** Kinetics and transient states of the reaction between a single SV and MPS. There are two energy barriers, the first one (0.51eV) is due to the S-H bond breaking, and the second one (0.22eV) is due to S-C bond breaking. (a)-(e) plots the initial, transient, and final states of the reaction. The SV in the initial state is illustrated by dashed line. The inset shows the chemical structure of MPS.

**Figure 2** High-resolution aberration-corrected TEM images of (a) as-exfoliated and (b) TS-treated monolayer MoS$_2$ sample, showing the significant reduction of SV by MPS treatment. The SVs are highlighted by red arrows. The overlaid blue and yellow symbols mark the position of Mo and S atoms respectively. The scale bars are 1nm. Detailed intensity profile analysis and histogram of SV density are shown in Supplementary Fig. 9.

**Figure 3** The effect of defect and interface engineering on monolayer MoS$_2$ charge transport. (a) Typical $\sigma$-$V_g$ characteristics for as-exfoliated (black), TS-treated (blue), and DS-treated (red) monolayer MoS$_2$ at $T$=300K. (b) $\mu$-$T$ characteristics for the three devices at $n$=7.1x10$^{12}$cm$^{-2}$. Solid lines are the best theoretical fittings. The dashed red line shows $T^{-0.72}$ scaling. (c) $\mu$-$n$ characteristics for the three devices at $T$=80K. Solid lines are the best theoretical fittings. (d)-(f) Arrhenius plot of $\sigma$ (symbols) and theoretical fittings (lines) for the as-exfoliated (d), TS-treated (e) and DS-treated (f) MoS$_2$. From top to bottom, $n$=7.0, 6.0, 5.0, and 4.0x10$^{12}$ cm$^{-2}$ respectively. The critical points of the MIT are highlighted by solid symbols in (e) and (f). Insets in (d)-(f) show the cartoon illustration of the corresponding MoS$_2$ samples undergone different treatments.

**Figure 4** Theory of charge transport in MoS$_2$. (a) Phase diagram of charge transport in monolayer MoS$_2$. The solid black line plots the calculated MIT critical points (using the parameters of DS-treated sample in Table 1) that separate the metallic and insulating regimes. The red symbols are experimental MIT points extracted from Fig. 3f. The lower left corner of the phase diagram illustrates the hopping transport regime (not calculated). (b) The solid black line is the calculated $n_0$ as a function of $N_{tr}$ using the parameters of DS-treated sample in Table 1. The MIT curves under different $N_{tr}$ are plotted in Supplementary Fig. 14. The blue and red symbols are experimental points from the TS-treated and DS-treated samples in Fig. 3 respectively. We use the highest $n$ that exhibit MIT as $n_0$. The horizontal dashed lines are MIT critical density estimated from Ref. 2 and 9 respectively. The intersections with the solid line represent the estimate of $N_{tr}$ in their devices (black arrows). (c) Theoretical calculation of $\mu$ as a function of $N_{tr}$ at $T$=300K without any CI or short-range scatterings (using the parameters of DS-treated sample in Table 1). From top to bottom, $n$=2.0, 1.2, 0.7 and 0.16x10$^{13}$ cm$^{-2}$ respectively. The phonon-limited value of 410cm$^2$V$^{-1}$s$^{-1}$ is recovered at $N_{tr}$=0.

**Table 1** The fitting parameters in our theoretical model for the three devices in Fig. 3.

**Figure 1**

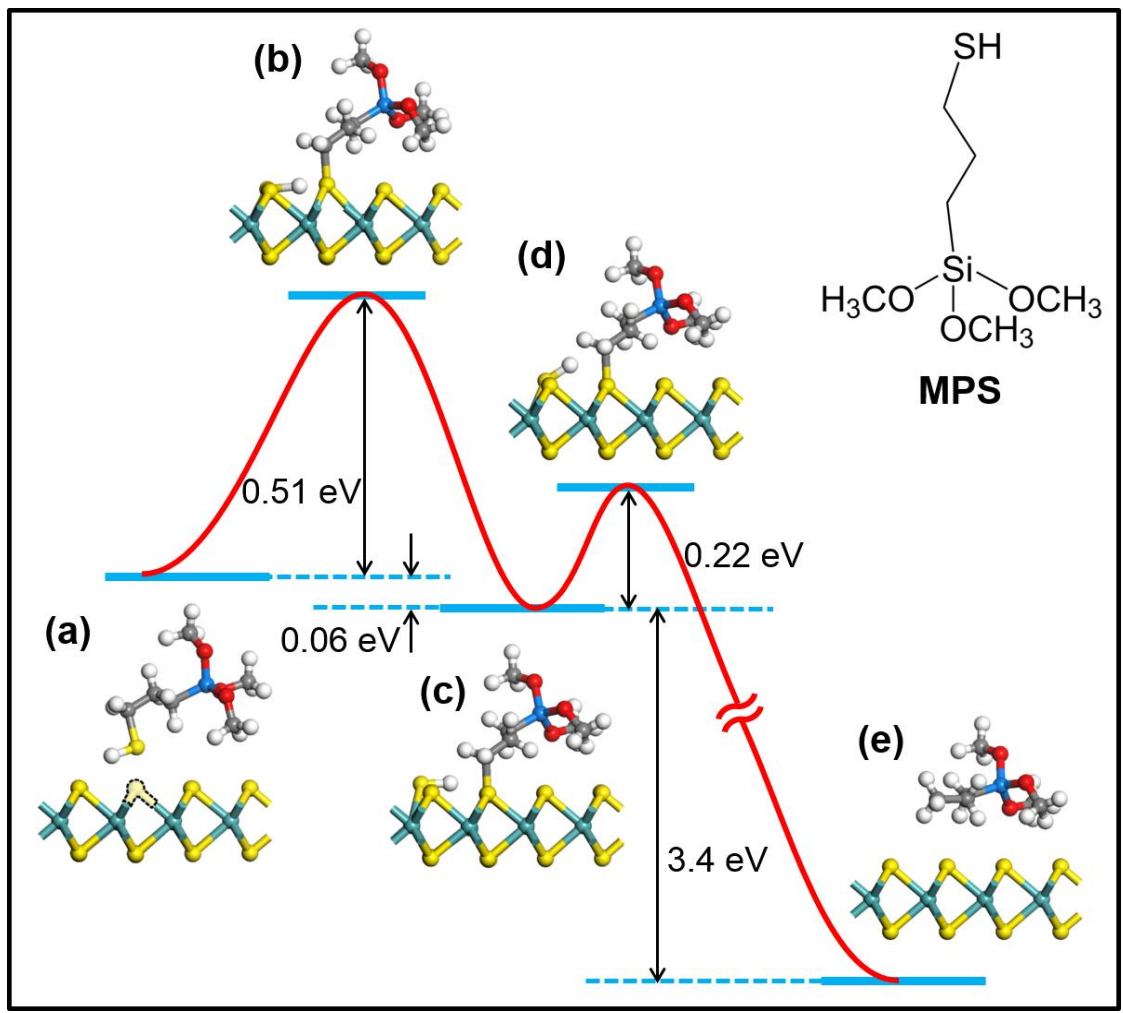

**Figure 2**

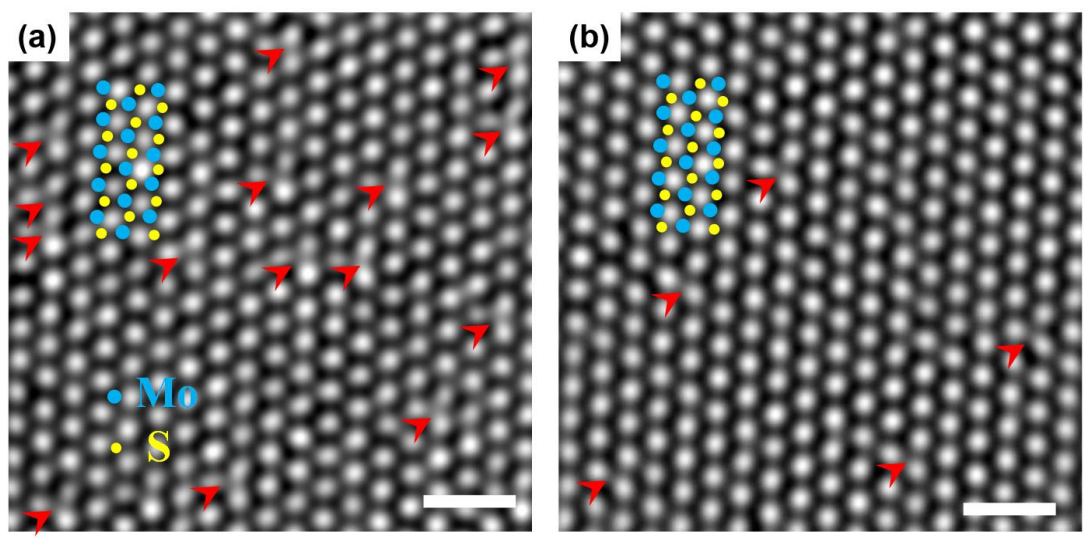

**Figure 3**

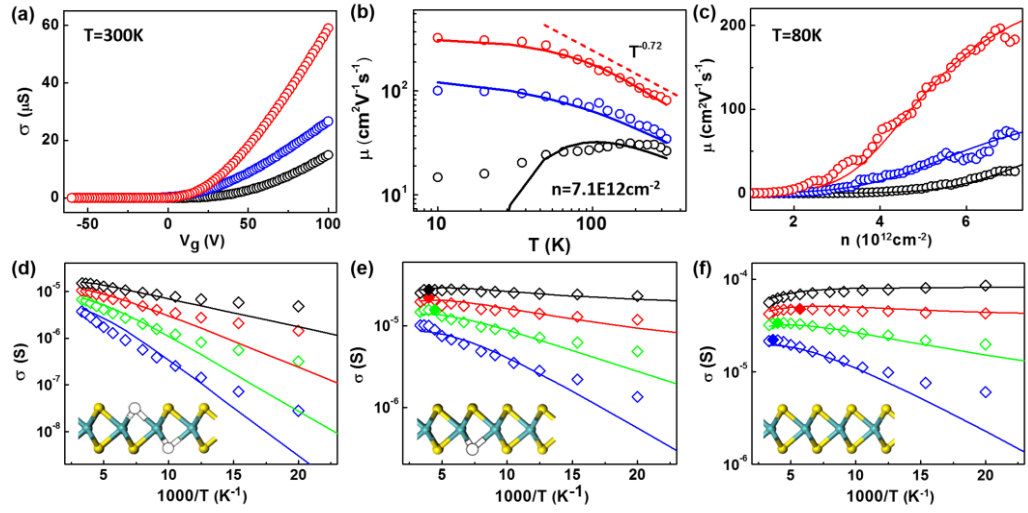

**Figure 4**

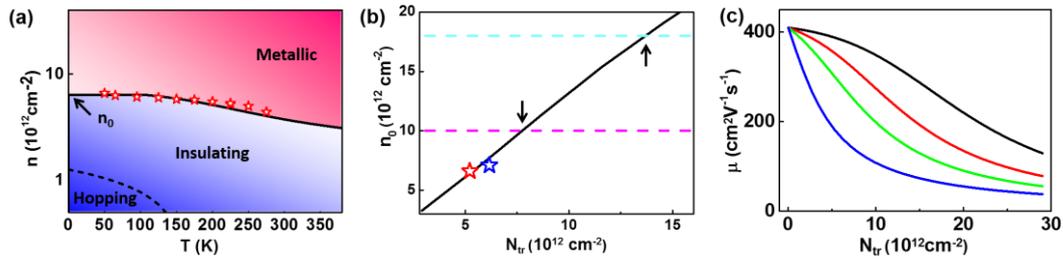

**Table 1**

|  | as-exfoliated | TS-treated | DS-treated |
|---|---|---|---|
| $N_i$ ($10^{12}cm^{-2}$) | 0.7 | 0.59 | 0.24 |
| $N_{tr}$ ($10^{12}cm^{-2}$) | 8.1 | 6.16 | 5.22 |
| $\Delta E_{tr}$ (meV) | 75 | 46 | 58 |
| $\mu_{sr}$ ($cm^2V^{-1}s^{-1}$) | 127 | 161 | 410 |
26

**Supplementary Information:**

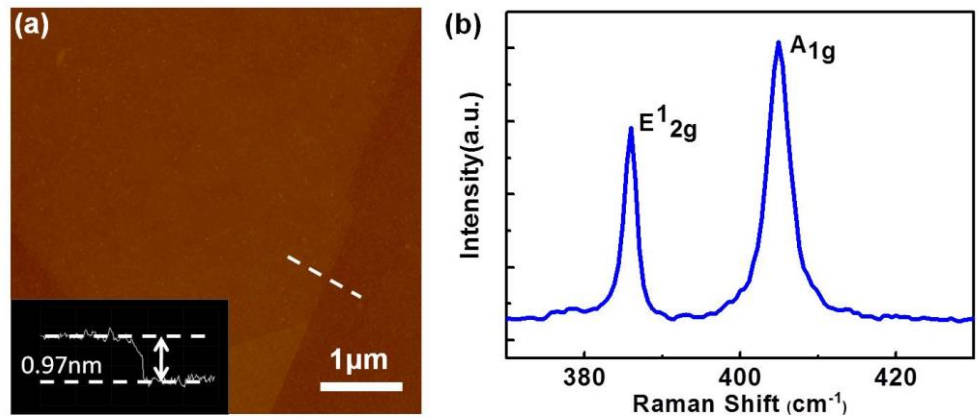

Supplementary Figure 1. Characterizations of monolayer $MoS_2$ studied in this work. We combine optical microscope, AFM and Raman spectroscopy to identify monolayer $MoS_2$ samples. (a) AFM image of a typical monolayer as-exfoliated $MoS_2$ sample on $SiO_2$/Si substrate. The height is ~1nm. (b) Raman spectrum of a typical monolayer $MoS_2$ sample. The position and relative intensity of the $E^1_{2g}$ mode (386 cm$^{-1}$) and $A_{1g}$ mode (405 cm$^{-1}$) confirm the monolayer nature[1].

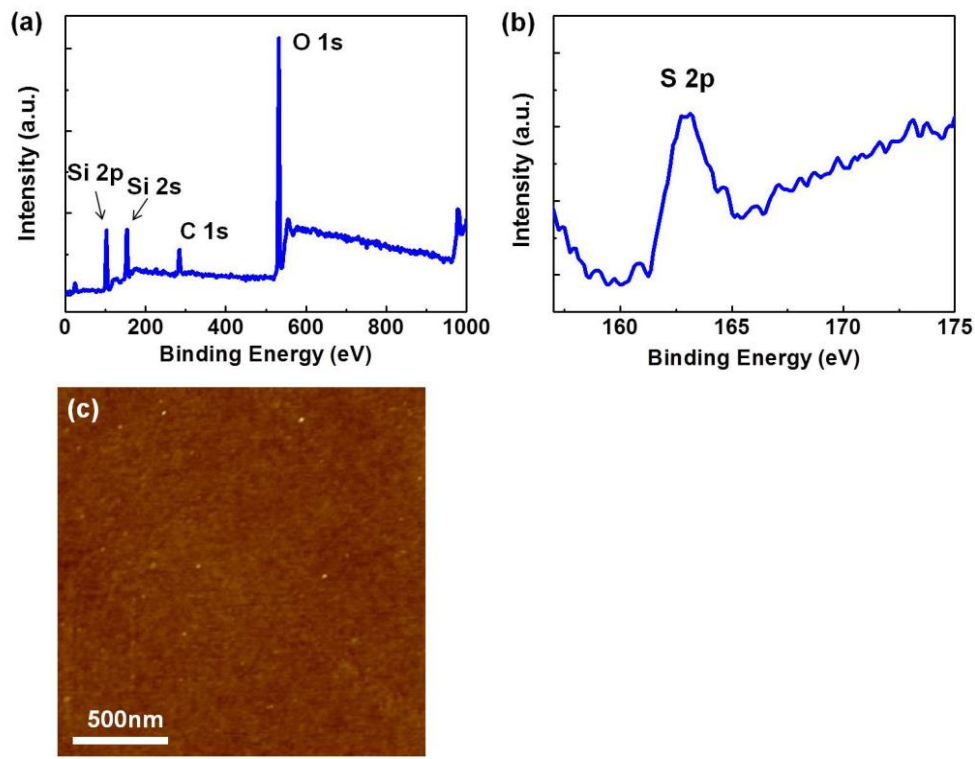

Supplementary Figure 2. Characterization of MPS grown on $SiO_2$ substrate. (a) X-ray photoelectron spectroscopy (XPS) survey scan of MPS SAM grown on 300nm $SiO_2$/Si substrate. The signal of Si, O, and C elements are most prominent. (b) High-resolution XPS scan near the S 2p region clearly showing the S signal from MPS. (c) AFM image of $SiO_2$ substrate after MPS SAM growth. The low roughness and good uniformity indicate the high quality of SAM growth.

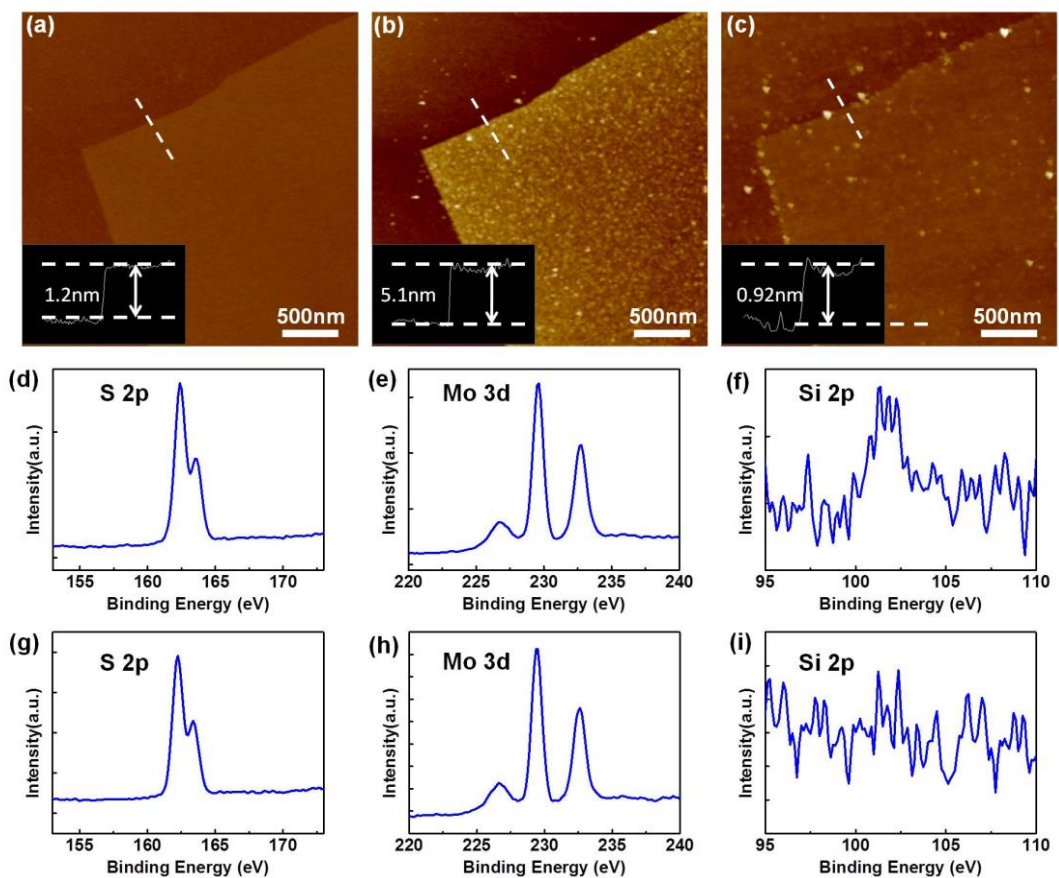

Supplementary Figure 3. Characterization of MoS$_2$ samples subject to MPS treatment. AFM images of the same monolayer MoS$_2$ sample (a) as-exfoliated on SiO$_2$/Si substrate, (b) after immersion in MPS solution and (c) after 350 ℃ annealing in H$_2$/Ar. The insets show the height of the sample at each stage. The apparent height of MoS$_2$ increases significantly after immersion in MPS, indicating that a thick layer of MPS is grown on top of the MoS$_2$. After the annealing step, the height of MoS$_2$ is restored, indicating that the extra MPS on MoS$_2$ is removed. (d)-(f) X-ray photoelectron spectroscopy (XPS) of MPS-treated MoS$_2$ and (g)-(i) as-exfoliated MoS$_2$. The existence of Si 2p peak in (f) clearly shows that the MPS grows on top of MoS$_2$.

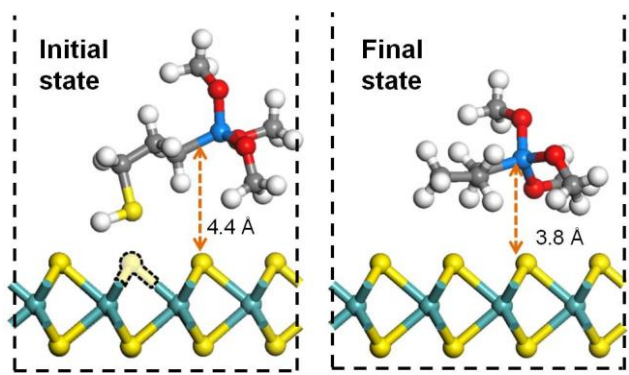

Supplementary Figure 4. Initial and final state of the chemical reaction in Fig. 1 of main text. The cyan, yellow, grey, white, red and blue balls denote Mo, S, C, H, O and Si atoms, respectively. The SV site is indicated by dashed line.

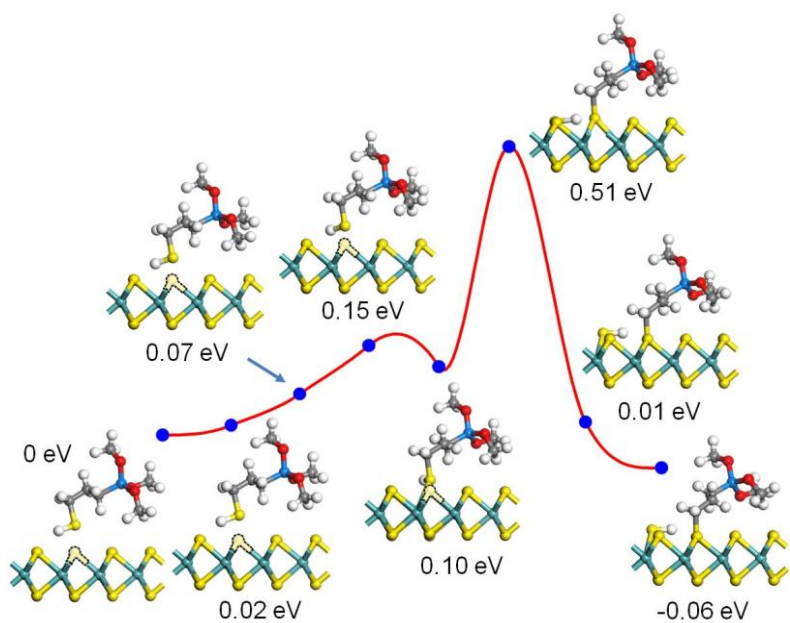

Supplementary Figure 5. Reaction kinetics and transient states of the S-H bond breaking step in the SV repair process.

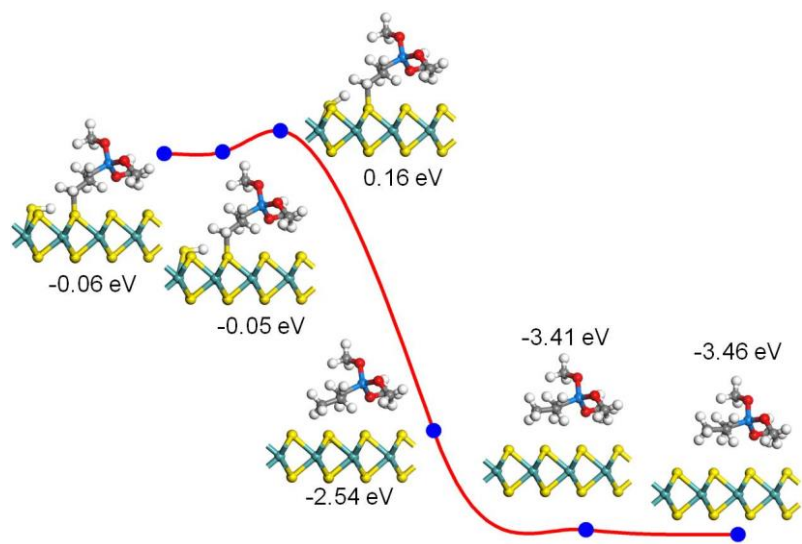

Supplementary Figure 6. Reaction kinetics and transient states of the S-C bond breaking step in the SV repair process.

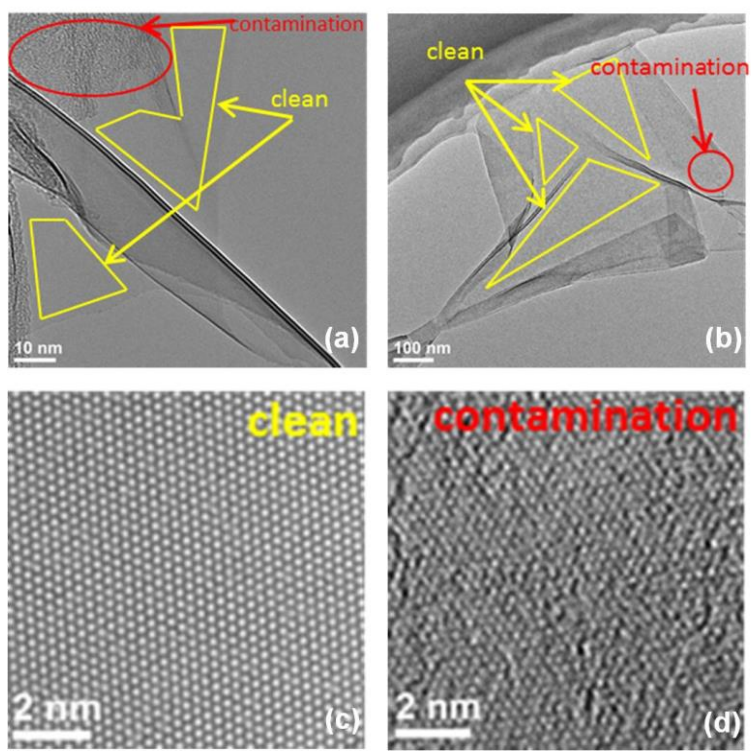

Supplementary Figure 7. (a) (b) Low-magnification TEM images of MoS$_2$ samples showing the areas with and without organic residue from the transfer process. (c) (d) High-resolution TEM images of clean area and contaminated area, respectively.

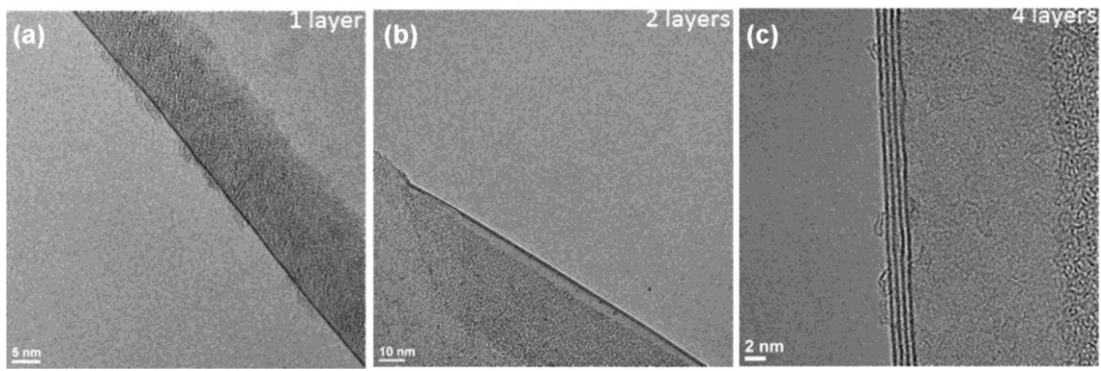

Supplementary Figure 8. Low-magnification TEM images of the edge of (a) monolayer (b) 2-layer and (c) 4-layer MoS$_2$ samples.

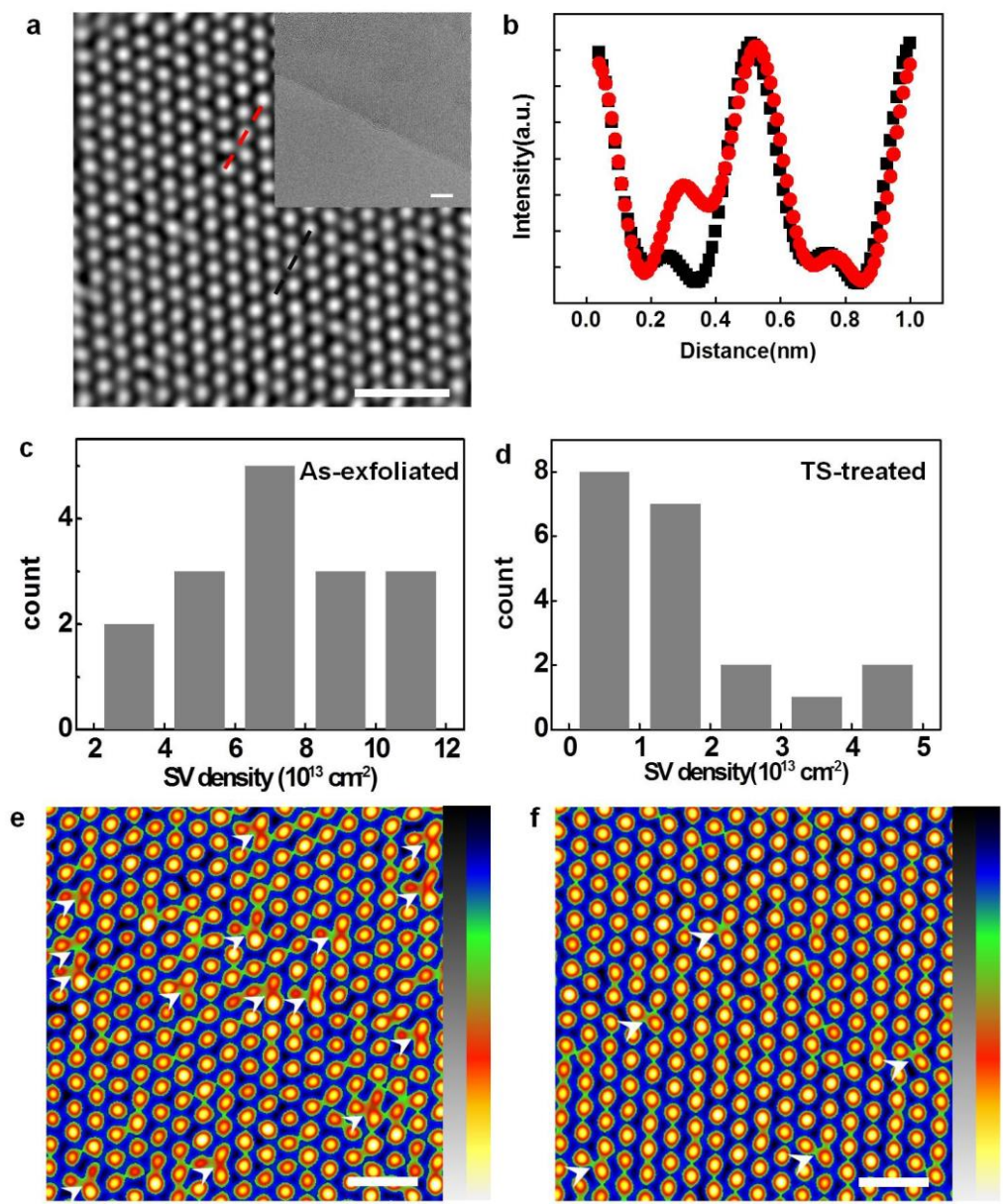

Supplementary Figure 9. (a) TEM image of the TS-treated monolayer $MoS_2$ in Fig. 2 of main text. Inset shows a low-magnification image of the $MoS_2$ sample edge confirming its monolayer nature. (b) Intensity profile along lattices with (red) and without (black) a SV. The profiles are taken at the dashed lines in (a). Compared to regular lattice, the SV shows brighter contrast, making it possible to identify them with high accuracy[2]. (c) and (d) are histogram of the density of SV in (c) as-exfoliated and (d) TS-treated monolayer $MoS_2$. The average density of SV is estimated to be $6.5 \times 10^{13} cm^{-2}$ and $1.6 \times 10^{13} cm^{-2}$, respectively. (e) and (f) are the same TEM images of Fig. 2 in main text. A band-pass filter to the Fast Fourier Transformation of the original images and false-colour reconstruction are applied to enhance the visibility of the SV (pointed by white arrows).

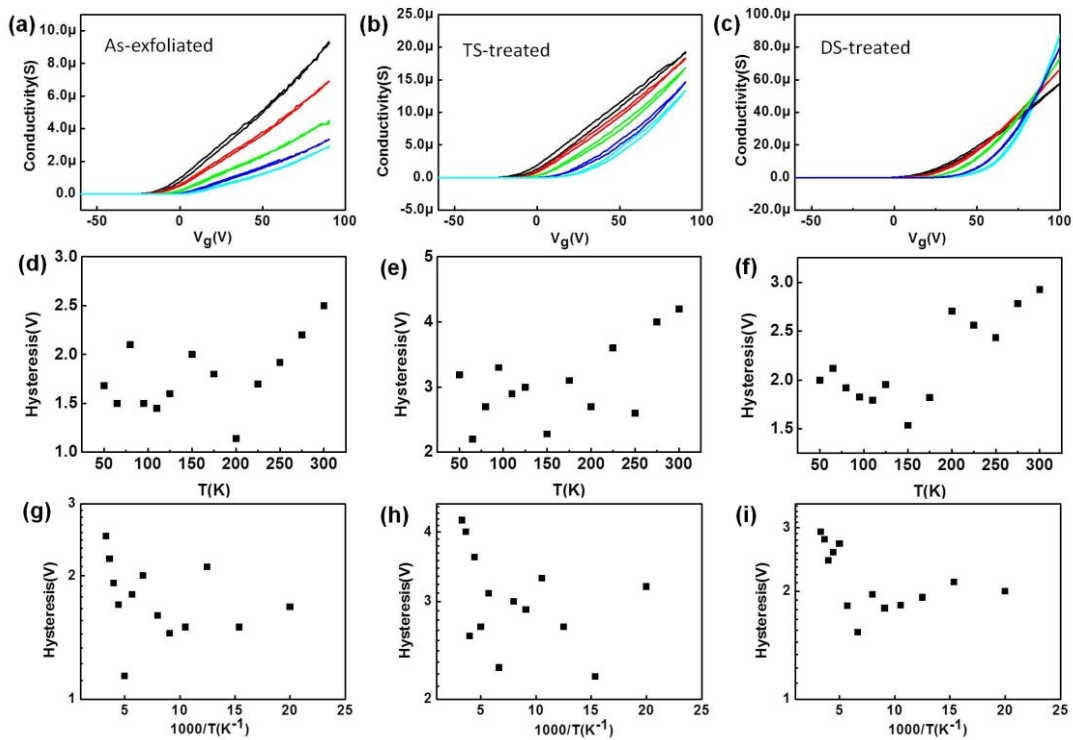

Supplementary Figure 10. Typical double-sweep transfer characteristics of (a) as-exfoliated, (b) TS-treated and (c) DS-treated monolayer MoS2 devices at different

temperatures. Black: 300K, red: 225K, green: 150K, blue: 80K, cyan: 50K. (d)-(f) are the hysteresis as a function of temperature for the three devices respectively. (g)-(i) are the Arrhenius plot of hysteresis for the three devices respectively.

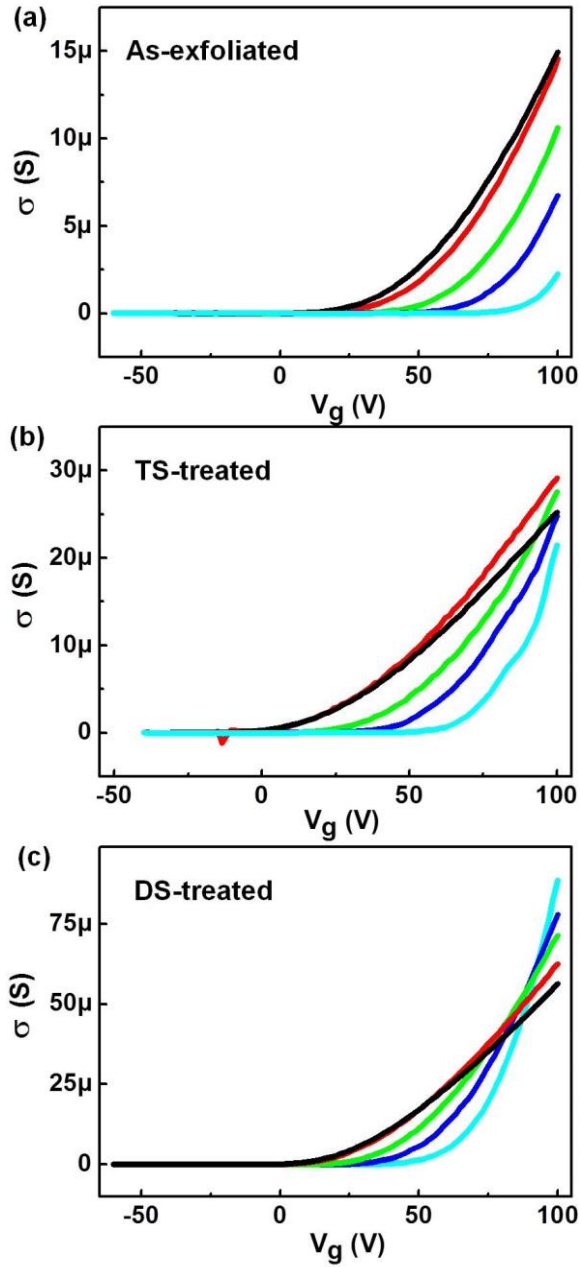

Supplementary Figure 11. Four-probe $\sigma$ vs $V_g$ characteristics for (a) as-exfoliated, (b) TS-treated, and (c) DS-treated monolayer $MoS_2$ in Fig. 3. In each panel, black, red, green, blue and cyan curve are taken at $T$=300K, 250K, 150K, 80K and 20K respectively.

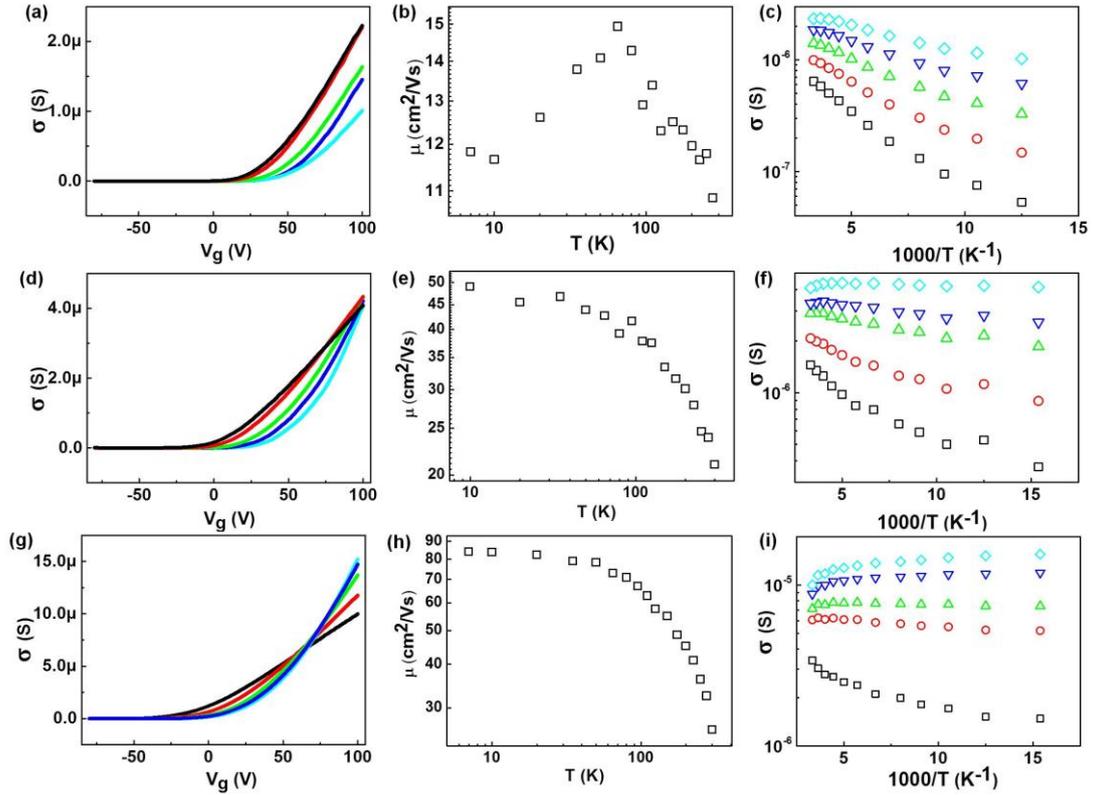

Supplementary Figure 12. Transport data of three additional monolayer MoS$_2$ devices in two-terminal geometry. (a), (d) and (g) are two-probe $\sigma$ vs $V_g$ characteristics for the as-exfoliated, TS-treated, and DS-treated monolayer MoS$_2$ sample respectively. In each panel, black, red, green, blue and cyan curve are taken at $T$=300K, 250K, 150K, 80K and 20K respectively. (b), (e) and (h) are $\mu$ vs $T$ for the as-exfoliated, TS-treated, and DS-treated monolayer MoS$_2$ sample at $n$=7.1x10$^{12}$cm$^{-2}$. (c), (f) and (i) are Arrhenius plots of $\sigma$ for the as-exfoliated, TS-treated, and DS-treated monolayer MoS$_2$ sample respectively. In each panel, from top to bottom, $n$=7.0, 6.0, 5.0, 4.0, 3.0x10$^{12}$cm$^{-2}$ respectively. The scaling of $\sigma$ and $\mu$ with temperature and carrier density are qualitatively the same as the four-terminal devices discussed in the main text. The absolute values of $\sigma$ and $\mu$ are smaller because of contact resistance.

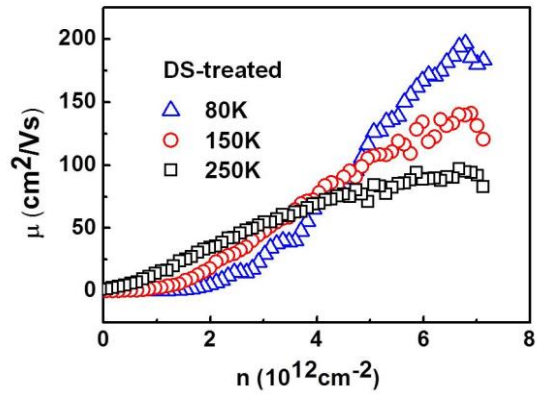

Supplementary Figure 13. $\mu$ vs $n$ for the DS-treated sample discussed in the main text at $T$=80K, 150K and 250K. The mobility edge clearly decreases with temperature.

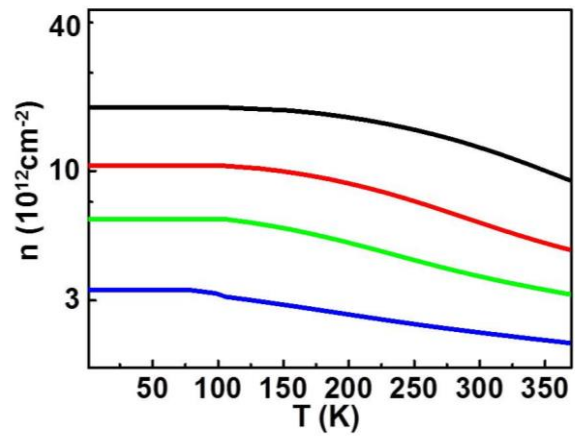

Supplementary Figure 14. Calculated MIT critical points under different $N_{tr}$, using the parameters of the DS-treated sample in Table 1 of main text. From top to bottom, $N_{tr}$=13.9, 8.12, 5.22 and 2.9x10$^{12}$cm$^{-2}$ respectively. From this figure, we can see that $n_0$ is roughly proportional to but slightly higher than $N_{tr}$.

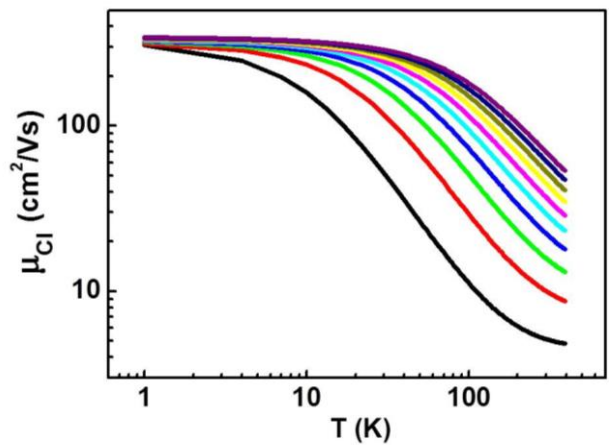

Supplementary Figure 15. Calculated $\mu_{CI}$ as a function of $T$ under $N_i=10^{12}$cm$^{-2}$. From bottom to top, $n$ is from $1.0\times10^{12}$ cm$^{-2}$ to $1.0\times10^{13}$ cm$^{-2}$ at a step of $1.0\times10^{12}$ cm$^{-2}$.

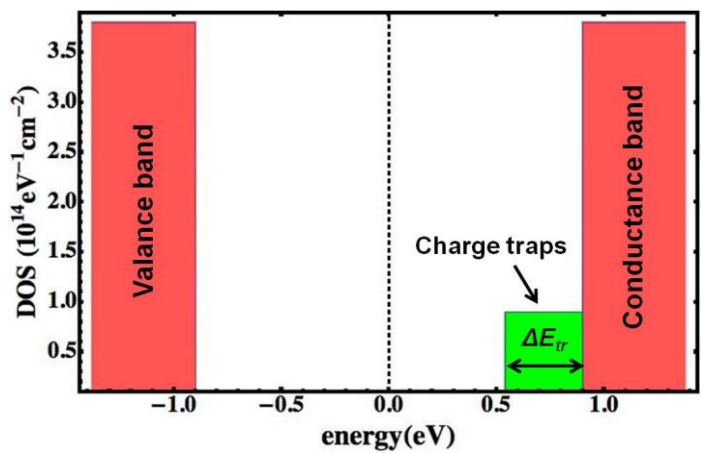

Supplementary Figure 16. Density of states of the valance band, conduction band and trap-induced impurity band used in our model.

Supplementary Note 1. After transfer to TEM grid, there are some organic residues on the MoS$_2$ samples. Usually the contaminations are not everywhere, but only on certain part of a sample. During TEM, we choose the area without contamination to take high-resolution images and analyze the defects. Supplementary Figure 7a and b show low-magnification TEM images of two MoS$_2$ samples, and we can clearly identify the clean and contaminated areas respectively. In addition, contaminated area looks very different from the clean area in high-resolution images (Supplementary Fig 7c, d), because the organic contaminations are long chained polymers. Therefore, the observed point defects are not organic contaminations but rather intrinsic defects from $MoS_2$. In addition, the number of layers can also be easily distinguished under TEM by looking at the sample edges (Supplementary Fig. 8). We always check the monolayer nature of $MoS_2$ samples before performing high-resolution TEM characterizations.